\documentclass[pre,preprint,showpacs]{revtex4}
\usepackage{amsmath} 
\usepackage{amsfonts}
\usepackage{amssymb}
\newcommand{\pII}[2]{\frac{\partial #1}{\partial #2}}
\newcommand{\pIII}[3]{\left(\frac{\partial #1}{\partial
      #2}\right)_{#3}}

\newcommand{\N}{\mathcal{N}}
\newcommand{\supar}[1]{^{\left(#1\right)}}
\newcommand{\dbar}{d\mkern-6mu\mathchar'26}
\begin{document}
\title{Non-equilibrium nanothermodynamics}
\author{J. Carrete}
\author{L. M. Varela \email{fmluis@usc.es}}
\author{L. J. Gallego}
\affiliation{Grupo de Nanomateriais e Materia Branda,
Departamento de F\'{i}sica da Materia Condensada,
Facultade de F\'{i}sica, Universidade de Santiago de Compostela,
E-15782 Santiago de Compostela, Spain}

\begin{abstract}
  Entropy production for a system outside the thermodynamic limit is
  formulated using Hill's nanothermodynamics, in which a macroscopic
  ensemble of such systems is considered. The external influence of the
  environment on the average nanosystem is connected to irreversible
  work with an explicit formula based on the Jarzynski equality. The entropy
  production retains its usual form as a sum of products of fluxes and
  forces and Onsager's symmetry principle is proven to hold for the
  average nanosystem, if it is assumed to be valid for the macroscopic ensemble, by
  two methods. The first one provides expressions that relate the
  coefficients of the two systems. The second gives a general
  condition for a system under an external force to preserve Onsager's
  symmetry.
\end{abstract}


\maketitle

Non-equilibrium thermodynamics tries to understand macroscopic systems
out of equilibrium, and particularly those in steady states, without having
to resort to the (unattainable) dynamic description of all their
microscopic degrees of freedom, using instead the same few macroscopic
variables as in the thermostatic case but allowing for situations
where they depend on time. Interest on this field can be traced back
to the works of Boltzmann on Thomson's hypothesis about the division
of a general process into a reversible and an irreversible part
\cite{DeGroot}. However, it was not until 1931 that a truly systematic
derivation of the thermodynamics of irreversible processes near
equilibrium was attained by Onsager \cite{Onsager1,Onsager2} and later
refined by Casimir \cite{Casimir}.

Entropy  production   is  perhaps   the  most  important   concept  in
non-equilibrium thermodynamics, totally  absent from thermostatics. It
is usual  to look at it  as a function  of two sets of  variables, the
thermodynamic     fluxes    $\left\{\phi_i\right\}$     and    forces
$\left\{F_i\right\}$, defined  in such a way that  this production can
be  expressed  as  a  sum  of  products  of  conjugates,  $\dot{\Delta
  S}=\sum\limits_i    F_i\phi_i$,   the    fluxes   being    zero   at
equilibrium.   This   expression  is   supplemented   by   a  set   of
phenomenological relations which gives the fluxes as functions of the
forces,  these relations being such that  the forces  cancel at
equilibrium.  It   is  an  experimental  fact  that   there  exists  a
neighbourhood of  equilibrium where the relations between  the two sets
of variables are linear, that is, $\phi_i=\sum\limits_j L_{ij}F_j$.

Onsager's main result \cite{Onsager1,Onsager2} is the symmetry of the
phenomenological coefficients $L_{ij}=L_{ji}$, proven on the basis of
two general hypothesis: regression of fluctuations and microscopic dynamic reversibility. Systems under
the effect of external magnetic fields or Coriolis forces
are exceptions already known to Onsager and later treated by Casimir
in Ref.~\onlinecite{Casimir}. More recently, it has been shown that the
second hypothesis can be dropped for certain models \cite{Gabrielli}.

In the last decades, interest in nanoscopic systems has led to put
them in the front line of science and technology. Important research from
the point of view of statistical mechanics has been done during the
last decade, leading to such notable results as the Jarzynski equality
\cite{Jarzynski} and the Evans \cite{Evans} and Crooks \cite{Crooks}
fluctuation theorems, which have been experimentally verified
\cite{Experimental}. A good overview of these topics can be found in
Ref.~\onlinecite{Bustamante}. Nevertheless, since the seminal work of Hill
\cite{Hill} in the early 60's, not much theoretical efforts were
dedicated to the strictly thermodynamic formalism in the nanoscale up
to the first years of the XXI century, when the same author revisited
his own work and renamed it as ``nanothermodynamics''
\cite{HillN1,HillN2}. Hill generalized the equations for open
systems introducing a term associated to the
number of small systems in a macroscopic ensemble of them that
explicitly takes into account the energetic contribution of surface
and edge effects, system rotation and translation, etc., usually
negligible for macroscopic systems.

In this context, it is of great importance to have a theoretical
framework for describing the operation of nanomachines. Therefore,
some thermodynamic results have been tentatively extented to systems
far from the thermodynamic limit. Particularly, regarding the theory
of non-equilibrium fluctuations (of interest for the development of
nanomotors), the validity of Onsager's reciprocal relations is
sometimes taken for granted \cite{nanomotors}. It is thus desirable to
put the non-equilibrium thermodynamics of small systems on firm
theoretical foundations. The formulation of such a non-equilibrium
nanothermodynamics is the main aim in this report. By analogy with
Hill's equilibrium theory, the number of nanosystems in an ensemble,
which can be modified by production, destruction and transport, is
introduced as a macroscopic variable that survives in the nanoscopic
description.

In order to study the thermostatics of a small system, Hill
\cite{Hill} started with a large number $\N$ of them, so that the
ensemble itself was a system in the thermodynamic limit.  The author
developed a theory suitable for measurement devices that interact with
many of the small systems in such a way that the relevant
thermodynamic quantities are not those of an individual nanosystem,
but their averages over a significant number of them, which in a
homogeneous system will be equal to their average over all of the
nanosystems. In other words, if each nanosystem is described by a set
of extensive variables
$\left\{X\supar{\alpha}\right\}_{\alpha=1}^{\nu}$, and the state of
the total system is characterized by
$\left\{X_t\supar{\alpha}\right\}_{\alpha=1}^{\nu}\cup\left\{\N\right\}$,
the quantities accessible to measurement are
$\left\{\bar{X}\supar{\alpha}:=\frac{X_t\supar{\alpha}}{\N}\right\}_{\alpha=1}^{\nu}$.
The entropy of a nanosystem can likewise be defined as
$S=\frac{S_t}{\N}$.

The total system obeys the usual set of thermostatic relations,
particularly the Gibbs and Euler equations in entropic form,
$dS_t=\sum\limits_{\alpha}
y\supar{\alpha}dX_t\supar{\alpha}-\frac{\varepsilon d\N}{T}$ and
$S_t=\sum\limits_{\alpha} y\supar{\alpha}X_t\supar{\alpha}-
\frac{\varepsilon\N}{T}$, with
$y\supar{\alpha}:=\pIII{S_t}{X_t\supar{\alpha}}{X_t\supar{\beta\ne\alpha},\N}$
and $\varepsilon:= -T\pIII{S_t}{\N}{X_t\supar{\alpha}}$. Obviously, in
the description of the overall system $\varepsilon$ is simply the
chemical potential associated with the number of nanosystems. However,
in the thermodynamics of small systems it is called the
\textit{subdivision potential}, a new variable with no analogue in
conventional thermodynamics.

To formulate the Euler equation for the nanosystems it is enough
to divide both terms in the Euler equation by $\N$. Taking into account
that $dS=\frac{1}{\N}\left[dS_t-Sd\N\right]$, the Gibbs equation can
also be formulated and, subtracting the two equations for
$dS$, an inhomogeneous pseudo-Gibbs-Duhem equation arises. Thus, the
thermodynamic equations for the average small system are:
\begin{subequations}\label{grp:small0}
\begin{align}
S&=\sum\limits_{\alpha}y\supar{\alpha}\bar{X}\supar{\alpha}-\frac{\varepsilon}{T}\label{eqn:small1}\\
dS&=\sum\limits_{\alpha}y\supar{\alpha}d\bar{X}\supar{\alpha}\label{eqn:small2}\\
-d\left(\frac{\varepsilon}{T}\right)&=-\sum\limits_{\alpha}\bar{X}\supar{\alpha}dy\supar{\alpha}\label{eqn:small3}.
\end{align}
\end{subequations}
\noindent Comparing equations \eqref{eqn:small2} and
\eqref{eqn:small1}, it becomes apparent that $S$ does not satisfy Euler's
theorem and thus it is not an homogeneous function of
$\left\{\bar{X}\supar{\alpha}\right\}$ in the nanothermodynamic
formalism. The thermostatics of a small system depends on its
environment through $\varepsilon$. Thus, a small system has more
degrees of freedom than its large counterpart. The additional
contribution to the entropy (or, equivalently, to the internal energy)
comes from the aforementioned interface, edge, rotation and traslation
effects, which must become negligible as the size of
the system is increased, if conventional thermodynamics is to be
recovered.

Recently, Ben-Amotz and Honig \cite{BenAmotz} have used the Jarzynski
equality to give a general expression, $dS=\frac{\left\langle\dbar
    W\right\rangle_{\chi\left(t\right)}}{T}+k_B\log\left\langle\exp\left(\frac{-\dbar
      W}{k_BT}\right)\right\rangle_{\chi\left(t\right)}$, for the
entropy production of a system under a time-dependent constraint
$\chi\left(t\right)$ in contact with a thermostat at the (possibly
also time-dependent) temperature $T$, averaged over the processes
compatible with that constrain ($\dbar W$ is the elementary work
associated to a particular process). If it is assumed that this kind
of operation amounts to an average over the ensemble of nanosystems
(a reasonable hypothesis since $\N$ is large) it is possible, using
\eqref{grp:small0}, to give an expression for the change in
$\varepsilon$ between times $0$ and $t_0$ during the process
determined by $\chi\left(t\right)$, suitable for measurement or simulation:
\begin{align}
&\Delta\varepsilon_{\chi\left(t\right)}=T\sum\limits_{\alpha}
\left.y\supar{\alpha}X\supar{\alpha}\right|_0^{t_0}\nonumber\\
&-\int\limits_{\chi\left(t\right)}\left[\left\langle
    \dbar W\right\rangle+k_BT\log\left\langle\exp\left(\frac{-\dbar
    W}{k_BT}\right)\right\rangle\right].
\label{eqn:benamotz2}
\end{align}

The central part of this report is devoted to the application of
Hill's course of reasoning to a system out of equilibrium in the
thermodynamic branch (linear regime). For simplicity, only the case
with two homogeneous macroscopic subsystems ($A$ and $B$) will be
considered. The results can be readily generalized to an arbitrary
number of partitions or even to a continuous distribution, as long as
large enough macroscopic differential volumes are taken in order to
assure that they contain sufficient numbers of nanosystems. The method
used is valid as long as differential calculus can describe the
changes in the variables of the nanosystems to a good approximation
(i.e. they are not too small).

Suppose that the systems are separated by a diathermic, permeable and
deformable wall and slightly out of equilibrium with each other. With
the total system $A\cup B$ completely isolated, the total deformation
variables
$\left\{X_t\supar{\alpha}=X_{At}\supar{\alpha}+X_{Bt}\supar{\alpha}\right\}_{\alpha=1}^{\nu}$
are conserved. $\N=\N_A+\N_B$, however, can vary since it is perfectly
conceivable that the nanosystems (e.g. micelles) could split or merge
even in a macroscopically isolated system. Conservation of $\N$ would
be a reasonable assumption in two opposite limits: static nanosystems
whose dissociation energy is so high that interactions with their
environment cannot split them, and highly dynamic nanosystems which
are continuously reorganizing, but in such a way that the fluctuations
in the total number of systems are small compared to the average
value. It is possible to choose the time derivatives of all extensive
thermodynamic variables as fluxes. Expanding $\Delta
S_t=S_t-S_{t,eq}= \Delta S_{At}+\Delta S_{Bt}$ to second order in
these coordinates:
\begin{align} \Delta
    S_t&=\sum\limits_{I\in\left\{A,B\right\}}\sum\limits_{\alpha}\Delta
    X_{It}\supar{\alpha}\left[y_I\supar{\alpha}
      +\frac{1}{2}\sum\limits_{\beta}\pII{y_I\supar{\alpha}}{X_{It}\supar{\beta}}\Delta
      X_{It}\supar{\beta}\right]+\nonumber\\
    &+\sum\limits_{I\in\left\{A,B\right\}}\Delta\N_I\left[-\frac{\varepsilon_I}{T}+\sum\limits_{\alpha}\pII{y_I\supar{\alpha}}{\N_I}\Delta
      X_{It}\supar{\alpha}\right]+\nonumber\\
    &+\frac{1}{2}\sum\limits_{I\in\left\{A,B\right\}}\frac{\partial^2S_{It}}{\partial\N_I^2}\left(\Delta\N_I\right)^2,\label{eqn:Groot1}
\end{align}
\noindent where $\Delta X$ denotes the deviation of $X$ from
its equilibrium value and all the derivatives are evaluated at
equilibrium. Note that the bars have been dropped for notational
simplicity. This implies that $y_A\supar{\alpha}=y_B\supar{\alpha}$
for all $\alpha$ and
$\varepsilon_A\Delta\N_A=-\varepsilon_B\Delta\N_B$ and, given the
conservation of $X\supar{\alpha}$, it follows that $\Delta
X_A\supar{\alpha}=-\Delta X_B\supar{\alpha}$. Taking this into account
and differentiating the previous expression with respect to time:
\begin{align}
\dot{\Delta
  S}_t&=\sum\limits_{\alpha}\left[\pII{y_A\supar{\alpha}}{\N_A}\Delta\N_A-\pII{y_B\supar{\alpha}}{\N_B}\Delta\N_B+\right.\nonumber\\
&\left.+\sum\limits_{\beta}\left(\pII{y_A\supar{\alpha}}{X_{At}\supar{\beta}}+\pII{y_B\supar{\alpha}}{X_{Bt}\supar{\beta}}\right)\Delta
    X_{At}\supar{\beta}\right]\dot{\Delta X}_{At}\supar{\alpha}+\nonumber\\
&+\left[\sum\limits_{\alpha}\pII{y_A\supar{\alpha}}{\N_A}\Delta
  X_{At}\supar{\alpha}+\frac{\partial^2S_{At}}{\partial\N_A^2}\Delta\N_A\right]\dot{\Delta\N}_A+\nonumber\\
&+\left[-\sum\limits_{\alpha}\pII{y_B\supar{\alpha}}{\N_B}\Delta
  X_{At}\supar{\alpha}+\frac{\partial^2S_{Bt}}{\partial\N_B^2}\Delta\N_A\right]\dot{\Delta\N}_B.\label{eqn:Groot2}
\end{align}
\noindent This expression has the form
of a sum of products of fluxes and forces, the terms inside square
brackets being the forces
$\left\{F\supar{\alpha}\right\}_{\alpha=1}^\nu$ and $F\supar{\N}$. To
translate this expression into the nanoscopic language, the following
equalities must be used:
\begin{subequations}\label{grp:trans}
\begin{align}
\dot{\Delta X}_{It}\supar{\alpha}&=\frac{d\Delta\left(\N_I X_I\supar{\alpha}\right)}{dt}
=\N_I\dot{\Delta X}_{It}\supar{\alpha}+\dot{\Delta
  \N}_IX_{It}\supar{\alpha}\label{eqn:trans:1}\\
\dot{\Delta S}&=\frac{\dot{\Delta S_t}}{\N}-S_t\frac{\dot{\Delta
    \N}_A+\dot{\Delta \N}_B}{\N^2}\label{eqn:trans:2}\\
\Delta X_{It}\supar{\alpha}&=\N_I\Delta
  X_I\supar{\alpha}+X_I\supar{\alpha}\Delta\N_I-\Delta X_I\supar{\alpha}\Delta\N_I,\label{eqn:trans:3}
\end{align}
\end{subequations}
\noindent giving the result
\begin{subequations}\label{grp:small}
\begin{align}
\dot{\Delta S}=&\sum\limits_{\alpha}F\supar{\alpha}\dot{\Delta
X}_A\supar{\alpha}+F_{\N_A}\dot{\Delta\N}_A+F_{\N_B}\dot{\Delta\N}_B\label{eqn:small:1},
\end{align}
\noindent with
\begin{align}
F\supar{\alpha}:=&F_t\supar{\alpha}\frac{\N_A}{\N}\label{eqn:small:2}\\
F_{\N_A}:=&\frac{F_{\N_At}+\sum\limits_{\alpha}F_t\supar{\alpha}X_A\supar{\alpha}}{\N}-\frac{S_t}{\N^2}\label{eqn:small:3}\\
F_{\N_B}:=&\frac{F_{\N_Bt}}{\N}-\frac{S_t}{\N^2},\label{eqn:small:4}
\end{align}
\end{subequations}
\noindent which means that the entropy production of the average small
system can also be written as a sum of products of fluxes and
forces. As mentioned previously, there exists a neighbourhood of equilibrium in which a set of
linear phenomenological relations between these variables holds. By means of equations
\eqref{grp:trans} and \eqref{grp:small}, the macro and
nanoscopic linear coefficients can be related:
\begin{subequations}\label{grp:relations}
\begin{align}
L_t\supar{\alpha\beta}&=\frac{\N_A}{\N}\left(\N_AL\supar{\alpha\beta}+X_A\supar{\alpha}L\supar{\N_A\beta}\right)+\nonumber\\
&+\frac{X_A\supar{\beta}}{\N}\left(\N_AL\supar{\alpha\N_A}+X_A\supar{\alpha}L\supar{\N_AN_A}\right)\label{eqn:relations:1}\\
L_t\supar{\alpha\N_I}&=\frac{1}{\N}\left(\N_AL\supar{\alpha\N_I}+X_A\supar{\alpha}L\supar{\N_A\N_I}\right)\label{eqn:relations:4}\\
L_t\supar{\N_I\alpha}&=\frac{1}{\N}\left(\N_AL\supar{\N_I\alpha}+X_A\supar{\alpha}L\supar{\N_I\N_A}\right)\label{eqn:relations:5}\\
L_t\supar{\N_I\N_J}&=\frac{L\supar{\N_I\N_J}}{\N};\,I,J\in\left\{A,B\right\}.\label{eqn:relations:6}
\end{align}
\end{subequations}
\noindent The macroscopic system satisfies Onsager's reciprocity by
hypothesis, i.e. $L_t\supar{\alpha\N_A}=L_t\supar{\N_A\alpha}$. It is
easy to see, starting with the last equations of the previous block
and progressively back-substituting, that the nanoscopic coefficients
are also symmetric in this situation. These proportionality relations
ensure that the second law of thermodynamics is obeyed by the average
systems (although it can be transitorily violated by a small system) a
topic also discused in Ref.~\onlinecite{BenAmotz}.

The quantity $S_t$ is indeterminate in one additive constant; therefore,
the component of the fluxes proportional to it, arising from the terms
in \eqref{eqn:small:3} and \eqref{eqn:small:4}, must be zero. This
last condition is equivalent to:
\begin{subequations}\label{grp:relations2}
\begin{align}
L\supar{\alpha\N_A}&=-L\supar{\alpha\N_B}\label{eqn:relations2:1}\\
L\supar{\N_A\N_A}&=L\supar{\N_B\N_B}=-L\supar{\N_A\N_B}\label{eqn:relations2:2}.
\end{align}
\end{subequations}

\noindent These equalities allow for further interpretation of
\eqref{grp:relations} using the change of variables
$\left(\N_A,\N_B\right)\rightarrow\left(\N,D\right)$, with
$D:=\N_A-\N_B$. The time derivative of $\N$ represents the creation of
nanosystems per unit time, while at fixed $\N$ the time derivative of
$D$ corresponds to the transport of small systems. The two possible causes
of variation of $\left(\N_A,\N_B\right)$ are thus decoupled by this
change. Furthermore, application of the Curie principle shows that
variation of $D$ can only be coupled to vectorial fluxes such as those
treated in this report, while variation of $\N$ can be coupled with
chemical (scalar) processes. In particular, they cannot be coupled
with each other, because of the different tensor rank of the forces
involved.

Equations \eqref{grp:relations2} are equivalent to stating that
$L\supar{\alpha\N}=L\supar{\alpha\N_A}$, $L\supar{DD}=L\supar{\N_A\N_A}$ and the
rest of the phenomenological coefficients involving $\N$ or $D$ are
zero. $L_t\supar{DD}$ is proportional, with a factor $\frac{1}{\N}$, to
its nanometric equivalent. The remaining coefficients are:
\begin{subequations}\label{grp:relationschanged}
\begin{align}
L_t\supar{\N\alpha}&=\frac{L_t\supar{\N_A\alpha}+L_t\supar{\N_B\alpha}}{2}=\frac{2\N}{\N+D}L\supar{\N\alpha}\label{eqn:relationschanged:1}\\
L_t\supar{\alpha D}&=\frac{L_t\supar{\N\N}X_A\supar{\alpha}}{\N}\label{eqn:relationschanged:2}.
\end{align}
\end{subequations}
\noindent The last equation can be used as a definition of $L\supar{\N\N}$
in the particular case in which $\N$ is fixed. This result means that the
contribution of a flux of nanosystems to that of an extensive
variable $X_{At}\supar{\alpha}$ is proportional to the amount
$X_A\supar{\alpha}$ of that variable that each nanosystem carries
with it in its transit from $A$ to $B$, as expected.

The above results can be reproduced by an alternate method closer to
the well-known macroscopic proof. This method rests on the hypothesis
of regression of fluctuations, which states that the reaction of the system
to a small deviation from equilibrium caused by an external force is
the same as if it was caused by a spontaneous fluctuation. Moreover,
if
$\left\{Z\supar{\alpha}\right\}_{\alpha=1}^{\nu+1}:=\left\{X_t\supar{\alpha}\right\}_{\alpha=1}^{\nu}\cup\left\{\N\right\}$,
dynamic reversibility can be expressed as $\left\langle\Delta
  Z\supar{\alpha}\left(t\right)\Delta
  Z\supar{\gamma}\left(t+\tau\right)\right\rangle=\left\langle \Delta
  Z\supar{\alpha}\left(t+\tau\right)\Delta
  Z\supar{\gamma}\left(t\right)\right\rangle$, which straightforwardly
gives $\left\langle\dot{\Delta Z}\supar{\alpha}\Delta
  Z\supar{\gamma}\right\rangle=\left\langle\Delta
  Z\supar{\alpha}\dot{\Delta
    Z}\supar{\gamma}\right\rangle$. Substituting here a linear
development for a system in the thermodynamic limit, analogous to that of previous sections, $\dot{\Delta
  Z}\supar{\alpha}=\sum\limits_{\beta}L_t\supar{\alpha\beta}z\supar{\beta}$,
with $z\supar{\beta}:=\frac{\partial S_t}{\partial Z\supar{\beta}}$,
and using the fact that $\left\langle
  Z\supar{\alpha}z\supar{\beta}\right\rangle=-k_B\delta\supar{\alpha\beta}$
(easy to prove for a macroscopic system, see for instance
Ref.~\onlinecite{DeGroot}) the symmetry of the phenomenological matrix
follows immediately. However, for the nanosystem, the temporal evolution of the internal
variables cannot be related only to their deviations from equilibrium,
adopting instead the more general form \cite{Landau}:

\begin{equation}
\dot{\Delta
  X}\supar{\alpha}=\sum\limits_{\beta}L\supar{\alpha\beta}\frac{\partial
S}{\partial X\supar{\beta}}+f\supar{\alpha}\left(t\right),
\label{eqn:fluct1}
\end{equation}

\noindent where $f\supar{\alpha}\left(t\right)$ represents a general
force acting on the system. From Eq. \eqref{eqn:trans:3}, $\Delta
X\supar{\alpha}=\frac{\Delta
  X_t\supar{\alpha}-X\supar{\alpha}_{t,eq}\Delta\N}{\N}$. Approximating
$\frac{1}{\N}\simeq\frac{1}{\N_{eq}}$ and taking into account that
derivatives at constant $\N$ are equal for the macroscopic and average
system, it follows directly that $\left\langle\Delta
  X\supar{\alpha}\frac{\partial S}{\partial
    X\supar{\beta}}\right\rangle=-\frac{k_B\delta\supar{\alpha\beta}}{\N}$. Thus,
temporal reversibility implies that:

\begin{equation}
L\supar{\alpha\beta}-L\supar{\beta\alpha}=\left\langle
  f\supar{\alpha}\Delta X\supar{\beta}-f\supar{\beta}\Delta X\supar{\alpha}\right\rangle.
\label{eqn:fluct3}
\end{equation}

\noindent This result represents a general condition for Onsager's
symmetry to hold when evolution of the system is conditioned by a general
external force. In this particular case, as it has previously been
shown, the external forces $f\supar{\alpha}$ are proportional to
$\Delta\N$ through constants $L\supar{\alpha\N}$, which makes the
right hand side of the previous equation trivially equal to zero, thus
proving the symmetry of the phenomenological submatrix involving only
the internal coordinates. This second method cannot prove anything
about the coefficients involving $\N$, an external variable. To show
their symmetry it is still necessary to relate them to their
macroscopical counterparts. This formulation, however, connects more
easily with the language of the fluctuation theorems.

New cross-transport phenomena associated with the new degrees of
freedom must appear in the nanosystems. As predicted by Eqs.
\eqref{grp:relationschanged}, if the number of small systems is kept
constant, these phenomena will consist simply in the exchange of
extensive variables transported along with the nanosystems. Dynamic
nanosystems, such as micelles, could be thus worthier of study. Since
Hill's equilibrium formalism has already been applied (and
successfully compared to experimental data) to nanostructures such as
nanosolids and nanowires \cite{nanosolid}, they could also be good
candidates to find these cross-phenomena; for instance, in electric or
thermal measurements.

\begin{acknowledgments}
  This work was supported by the Spanish Ministry of Education and
  Science in conjunction with the European Regional Development Fund
  (Grants Nos. FIS2005-04239 and FIS2007-66823-C02-02). J. Carrete
  wishes to thank the financial support of the Direcci\'{o}n Xeral de
  Ordenaci\'{o}n e Calidade do Sistema Universitario de Galicia, da
  Conseller\'{i}a de Educaci\'{o}n e Ordenaci\'{o}n
  Universitaria-Xunta de Galicia.
\end{acknowledgments}

\end{document}